\documentclass[a4paper,oneside]{saip}
\usepackage{amssymb}
\usepackage{mathrsfs}
\usepackage{upgreek}
\usepackage{multicol}
\usepackage{comment}

\begin{document}

\title{Configurational Temperature in the 3D XY Model}

%
%
\author{Kutloano Nkojoana\, \orcidlink{0000-0002-7788-3564}$^{1}$ and Anosh Joseph \,\orcidlink{0000-0003-4288-8207}$^{1}$}

\affil{$^{1}$\,National Institute for Theoretical and Computational Sciences,
School of Physics, and Mandelstam Institute for Theoretical Physics, University of the Witwatersrand, Johannesburg, Wits 2050, South Africa}

\email{\url{kutloano.nkojoana@gmail.com} \, \url{anosh.joseph@wits.ac.za}}

%

\begin{abstract}

We investigate the configurational temperature estimator as a diagnostic tool for Monte Carlo and Langevin simulations of the three-dimensional XY model with an imaginary chemical potential. This estimator depends only on the field configurations. It provides a stringent internal consistency check for numerical sampling algorithms. We perform simulations using both real Langevin dynamics and the Metropolis Monte Carlo algorithm on an $8^3$ lattice across a range of coupling values, $\beta = 0.2 - 0.7$. Our results for the action density are in excellent agreement with strong-coupling expansion predictions at small $\beta$, providing an important validation of both simulation approaches. The measured value of the configurational temperature estimator shows good agreement with its expected value of unity in the symmetric phase. However, systematic deviations appear in the ordered phase. We attribute these deviations primarily to finite-size effects and discretization artifacts associated with the relatively small lattice volume. Our results demonstrate that the configurational temperature estimator provides a valuable diagnostic for assessing thermalization and algorithmic correctness in lattice field theory simulations. Such diagnostics are particularly important in preparation for studies at real chemical potential, where sign problems arise and conventional validation methods become less reliable.

\end{abstract} 

\section{Introduction}
\label{sec:intro}

Understanding the properties of strongly interacting matter at finite baryon density remains one of the major challenges in lattice field theory \cite{Meyer-Ortmanns:1996ioo}. 
Although lattice Quantum Chromodynamics (QCD) has achieved remarkable success at vanishing chemical potential, conventional importance-sampling methods become ineffective at finite density because the Euclidean action generally acquires a complex phase \cite{Nagata:2021ugx}. 
This sign problem prevents a probabilistic interpretation of the Boltzmann weight and severely limits first-principles numerical studies of the QCD phase diagram. 

Among the approaches developed to overcome the sign problem, complex Langevin dynamics has emerged as a promising alternative since it does not rely on importance sampling. 
Instead, field configurations are generated through a stochastic evolution in a complexified field space. 
While this method has been successfully applied to a variety of theories, its convergence to the correct equilibrium distribution is not guaranteed. 
Consequently, reliable diagnostics capable of assessing the correctness of numerical simulations remain an important area of current research. 

One such diagnostic is the \emph{configurational temperature} \cite{Rugh:1997, Butler:1998, Jepps:2000, Dhindsa:2025xfv, Joseph:2025xbn, Longia:2026doi, Joseph:2026xti, Joseph:2026hom, Joseph:2026rwh}. 
This observable is constructed from the derivatives of the Euclidean action and, therefore, depends only on the sampled field configurations, in contrast to conventional thermodynamic observables. 
For correctly sampled equilibrium ensembles, it is expected to reproduce a known reference value, making it a useful internal consistency check of numerical algorithms. 
Since it does not require comparison with independent analytical predictions, the configurational temperature is particularly attractive for simulations of systems where conventional validation methods are unavailable. 

As a first step towards applying this diagnostic to theories with complex actions, we investigate its behavior in the three-dimensional XY model at an imaginary chemical potential \cite{Banerjee:2010kc}. 
In this regime, we have a real Euclidean action, allowing both Metropolis Monte Carlo and real Langevin simulations to be performed under controlled conditions. 
The XY model, therefore, provides an ideal benchmark for testing configurational diagnostics before extending them to theories that suffer from a genuine sign problem. 

In this work, we perform simulations on an $8^3$ lattice over a range of couplings and imaginary chemical potentials. 
We first validate our numerical implementation by comparing measurements of the action density obtained using Metropolis and Langevin simulations with the predictions of the strong-coupling expansion. 
We then investigate the behavior of the configurational-temperature estimator across the phase diagram and compare the results obtained using the two simulation algorithms. 
Our study establishes a benchmark for future applications of configurational diagnostics to complex Langevin simulations at real chemical potential. 

\section{Three-Dimensional XY Model and Configurational Temperature} 
\label{sec:model} 

\subsection{Three-Dimensional XY Model} 

The Euclidean action of the three-dimensional XY model on a lattice is 
\begin{equation} 
S = - \beta \sum_x \sum_{\nu = 0}^2 \cos \left( \phi_x - \phi_{x + \hat{\nu}} \right), 
\label{eq:xy_action} 
\end{equation} 
where $\phi_x \in [0, 2 \pi)$ is an angular field defined on each lattice site, $\beta$ denotes the lattice coupling, and $\hat{\nu}$ labels the three lattice directions. 
Throughout this work, we consider an $N_s^2\times N_\tau$ lattice with periodic boundary conditions. 

A chemical potential is introduced by modifying the temporal hopping terms, 
\begin{equation} 
S = - \beta \sum_x \sum_{\nu = 0}^2 \cos \left( \phi_x - \phi_{x + \hat{\nu}} - i \mu \delta_{\nu, 0} \right), 
\label{eq:xy_mu_action} 
\end{equation} 
where $\delta_{\nu, 0}$ ensures that the chemical potential couples only to the temporal direction. 

For a real chemical potential, the action becomes complex, leading to the sign problem and preventing conventional importance-sampling Monte Carlo simulations. 
In this work, we instead consider an imaginary chemical potential, 
\begin{equation} 
\mu = i \mu_I, 
\end{equation} 
for which the action remains real. 
This allows us to perform both Metropolis Monte Carlo and real Langevin simulations and to compare the two algorithms under controlled conditions. 

\subsection{Configurational Temperature} 

The configurational temperature provides an internal diagnostic of equilibrium sampling using only derivatives of the Euclidean action. 
For a lattice action $S(\phi)$, the configurational-temperature estimator is defined by 
\begin{equation} 
\hat{\beta} = \frac{\langle \nabla^2 S \rangle}{\langle |\nabla S|^2 \rangle}.
\label{eq:beta_general} 
\end{equation} 
Here, $\nabla S$ and $\nabla^2S$ denote the gradient and Laplacian of the action with respect to all field variables. 
For correctly sampled equilibrium configurations, the estimator is expected to reproduce its equilibrium value and therefore serves as a useful consistency check of the simulation. 

For the XY model, the first derivative of the action is 
\begin{equation} 
\frac{\partial S}{\partial\phi_x} = \beta \sum_{\nu = 0}^2 \left[ \sin \left( \phi_x - \phi_{x + \hat{\nu}} - i \mu \delta_{\nu, 0} \right) - \sin \left( \phi_{x - \hat{\nu}} - \phi_x - i \mu\delta_{\nu, 0} \right) \right], 
\label{eq:gradient} 
\end{equation} 
while the corresponding second derivative is 
\begin{equation} 
\frac{\partial^2S}{\partial\phi_x^2} = \beta \sum_{\nu = 0}^2 \left[ \cos \left( \phi_x - \phi_{x + \hat{\nu}} - i \mu \delta_{\nu, 0} \right) + \cos \left( \phi_{x - \hat{\nu}} - \phi_x - i \mu \delta_{\nu, 0} \right) \right]. 
\end{equation} 

For every measured configuration, we compute 
\begin{align} 
g^2 &= \frac{1}{\Omega} \sum_{i = 1}^\Omega \left( \frac{\partial S}{\partial \phi_i} \right)^2, \\ 
H &= \frac{1}{\Omega} \sum_{i = 1}^\Omega \frac{\partial^2S}{\partial \phi_i^2}, 
\end{align} 
where $\Omega = N_s^2 N_\tau$ is the lattice volume. 
The configurational-temperature estimator for an individual configuration is then 
\begin{equation} 
\hat{\beta} = \frac{H}{g^2}, 
\label{eq:beta_single} 
\end{equation} 
and the measured configurational temperature is obtained from the ensemble average, 
\begin{equation} 
\beta_M = \left\langle \hat{\beta} \right\rangle. 
\label{eq:beta_average} 
\end{equation} 

\section{Numerical Simulations} 
\label{sec:numerical} 

We performed simulations of the three-dimensional XY model on an $8^3$ lattice with periodic boundary conditions. 
The lattice coupling was varied over the range 
\[ 
\beta = 0.2, \; 0.3, \; \ldots, \; 0.7.
\] 
The chemical potential was chosen to satisfy 
\[ 
\mu^2 = 0, \, - 0.05, \, - 0.10, \, - 0.15, \, - 0.20, 
\] 
corresponding to purely imaginary chemical potentials. 

To benchmark the configurational-temperature estimator, we generated the equilibrium configurations using both the Metropolis Monte Carlo algorithm and real Langevin dynamics. 
The action remains real at an imaginary chemical potential, and thus, both algorithms sample the same equilibrium distribution. 
Therefore, they provide an independent consistency check of the numerical implementation. 

We discarded the first $10^5$ updates for each parameter set for thermalization. 
The measurements were then subsequently accumulated over $5 \times 10^5$ production updates. 
To reduce autocorrelation, the observables were recorded every $100$ updates. 
The Langevin simulations employed an adaptive integration step size to maintain numerical stability throughout the evolution. 
Statistical uncertainties were estimated using the jackknife method. 

Two observables were measured throughout this study. 
The first is the action density, 
\begin{equation} 
\frac{\langle S \rangle}{\Omega}, 
\end{equation} 
which provides a conventional benchmark for validating the numerical simulations. 
At weak coupling, the results can be compared with the analytical prediction from the strong-coupling expansion, 
\begin{equation} 
\frac{\langle S\rangle}{\Omega} = - \frac{3}{2} \beta^2 - \frac{21}{16} \beta^4 + \mathcal{O}(\beta^6).
\label{eq:strong_coupling} 
\end{equation} 

The principal quantity of interest is the configurational-temperature estimator defined in the previous section. 
For correctly sampled equilibrium configurations, one expects 
\begin{equation} 
\beta_M \simeq 1, 
\label{eq:beta_expected} 
\end{equation} 
up to statistical uncertainties together with finite-volume and discretization effects. 
In the following section, we compare measurements obtained from the two simulation algorithms and investigate the behavior of the estimator across the parameter space. 

\section{Results} 
\label{sec:results} 

\subsection{Validation of the Numerical Simulations} 
\label{subsec:validation} 

Before investigating the configurational-temperature estimator, it is important to verify that both simulation algorithms correctly reproduce known equilibrium properties of the three-dimensional XY model. 
For this purpose, we compare measurements of the action density obtained from Metropolis Monte Carlo and real Langevin simulations with the predictions of the strong-coupling expansion. 

\begin{figure}[htbp] 
\centering 
\includegraphics[width=2.0in]{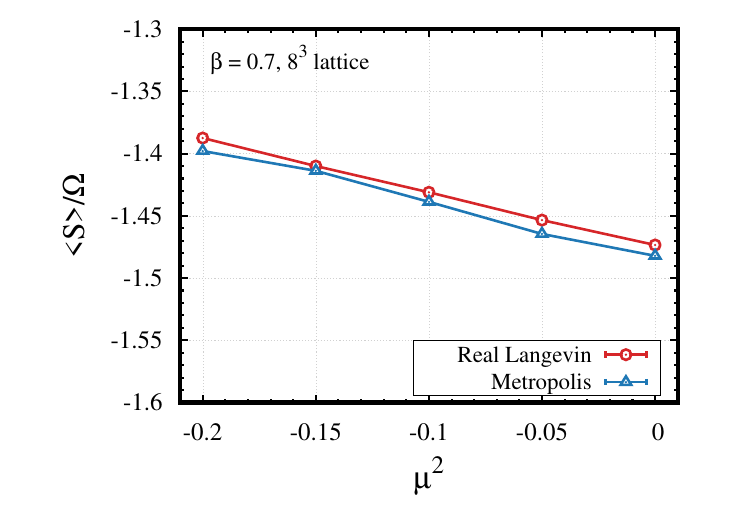} 
\includegraphics[width=2.0in]{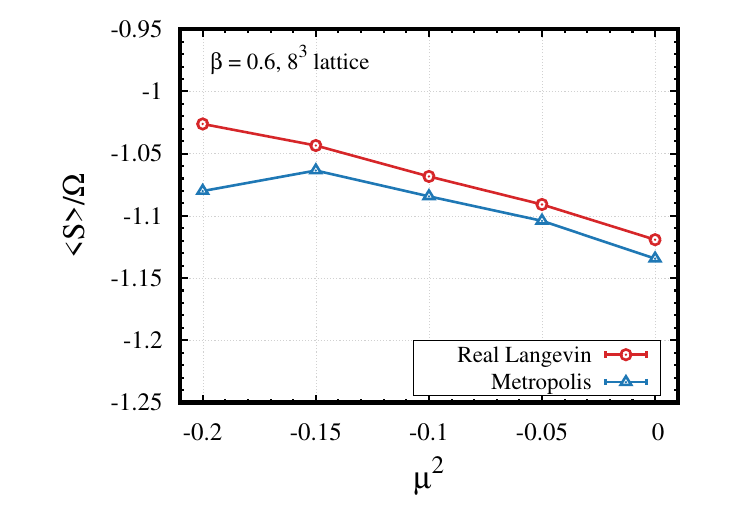} 
\includegraphics[width=2.0in]{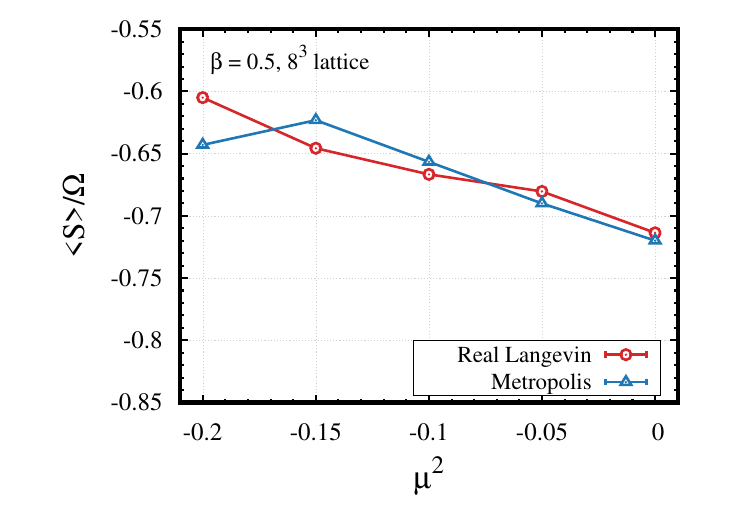} 
\includegraphics[width=2.0in]{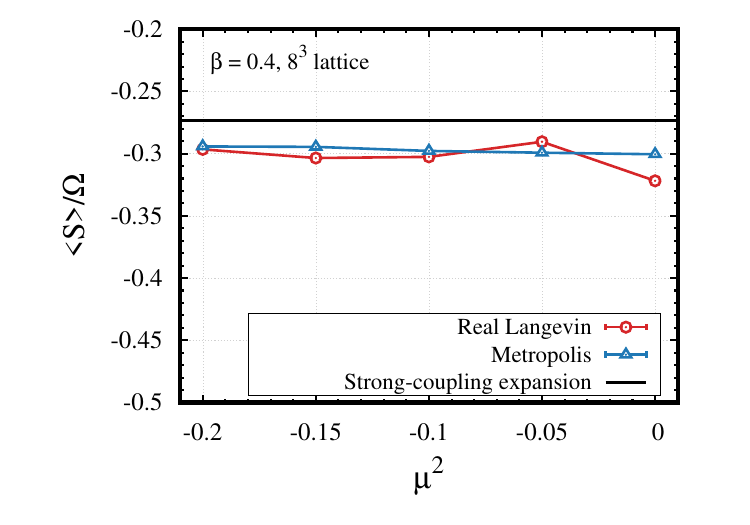} 
\includegraphics[width=2.0in]{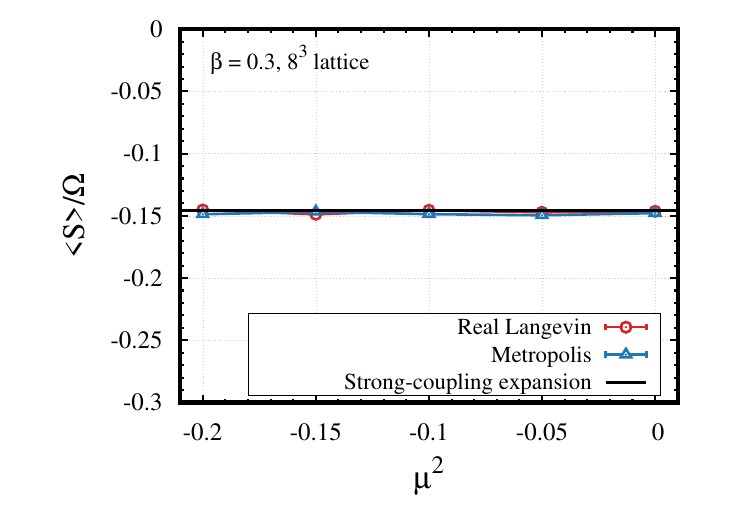} 
\includegraphics[width=2.0in]{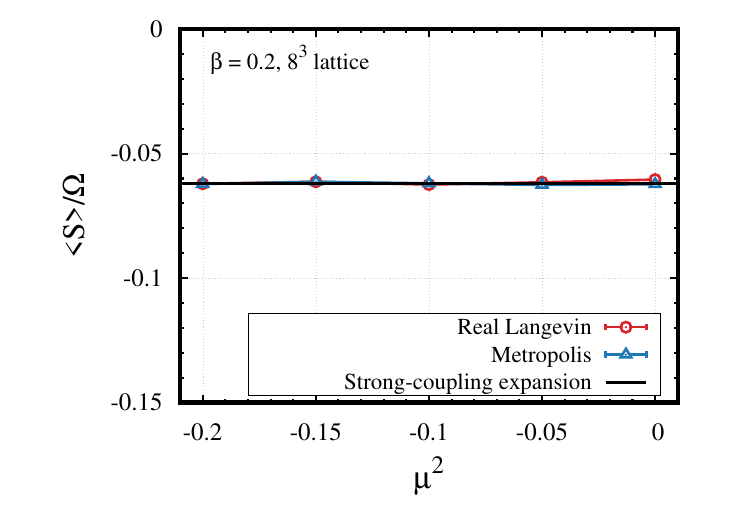} 
\caption{Action density $\langle S \rangle / \Omega$ as a function of $\mu^2$ on an $8^3$ lattice for several values of $\beta$. 
Results from real Langevin dynamics and Metropolis Monte Carlo are compared. 
The strong-coupling predictions are shown for the weak-coupling cases.} 
\label{fig:action_density} 
\end{figure} 

Figure~\ref{fig:action_density} shows the action density against the imaginary chemical potential for couplings in the range $\beta = 0.2$ -- $0.7$. 
Results obtained using the two simulation algorithms are displayed together. 
At weak coupling, the numerical measurements agree well with the strong-coupling expansion, providing an important validation of the numerical implementation. 
For example, at $\mu^2 = 0$ the real Langevin simulations yield 
\[ 
\frac{\langle S\rangle}{\Omega} = - 0.0609(2) 
\] 
for $\beta = 0.2$ and 
\[ 
\frac{\langle S\rangle}{\Omega} = - 0.1471(3) 
\] 
for $\beta = 0.3$, in good agreement with the corresponding strong-coupling predictions of $- 0.0621$ and $- 0.1450$. 

Across the full range of couplings we simulated, Metropolis Monte Carlo and real Langevin dynamics exhibit the same qualitative dependence on the chemical potential. 
Small quantitative differences become visible at intermediate and stronger couplings, where higher-order corrections, finite-volume effects, and statistical fluctuations are expected to become more significant. 
Importantly, these differences do not display any systematic trend indicative of numerical instabilities or incorrect sampling. 

The close agreement between the two independent simulation algorithms, together with the consistency with the strong-coupling expansion at weak coupling, provides strong evidence that the equilibrium ensembles are sampled correctly. 
Having established this benchmark, we now turn to the configurational-temperature estimator, which constitutes the principal focus of the present work. 

\subsection{Configurational-Temperature Estimator} 
\label{subsec:configtemp} 

\begin{figure}[htbp] 
\centering 
\includegraphics[width=2.0in]{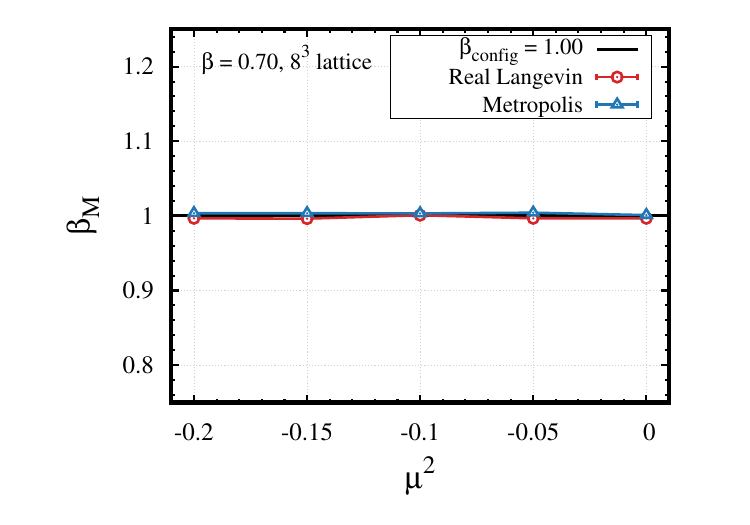} 
\includegraphics[width=2.0in]{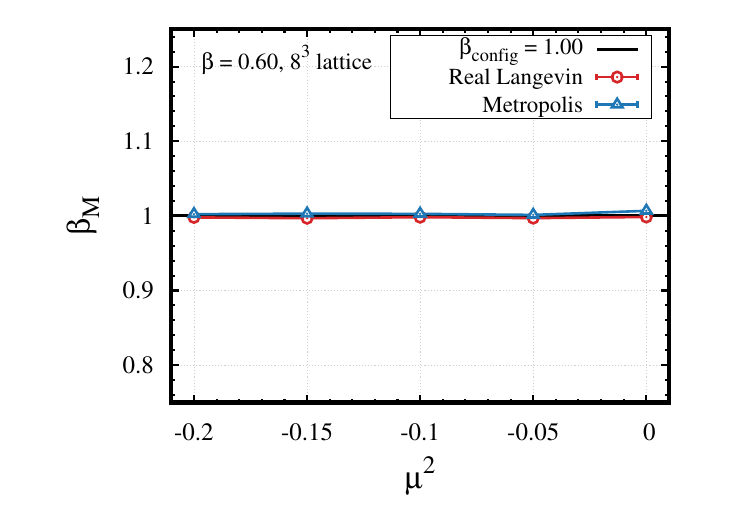} 
\includegraphics[width=2.0in]{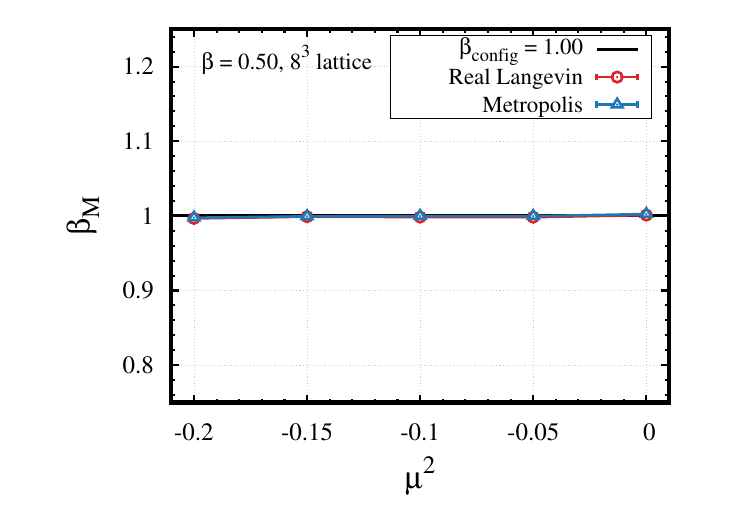} 
\includegraphics[width=2.0in]{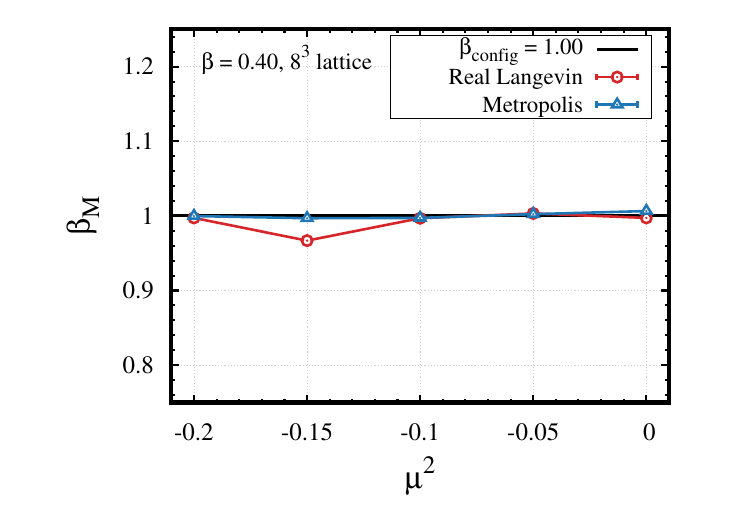} 
\includegraphics[width=2.0in]{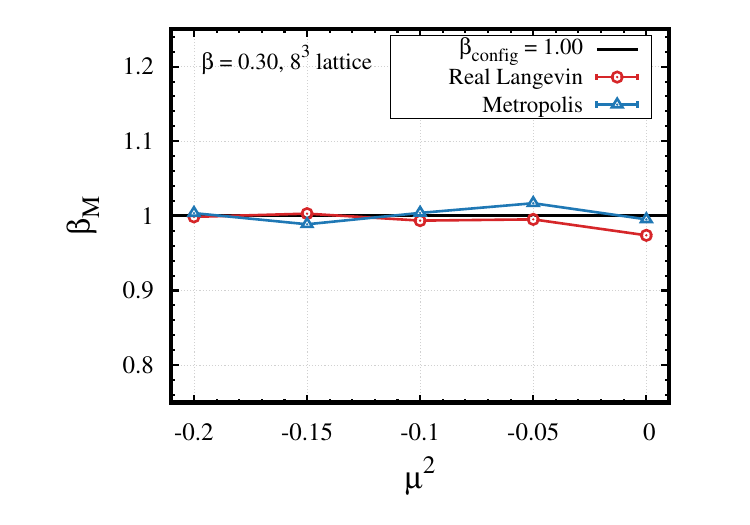} 
\includegraphics[width=2.0in]{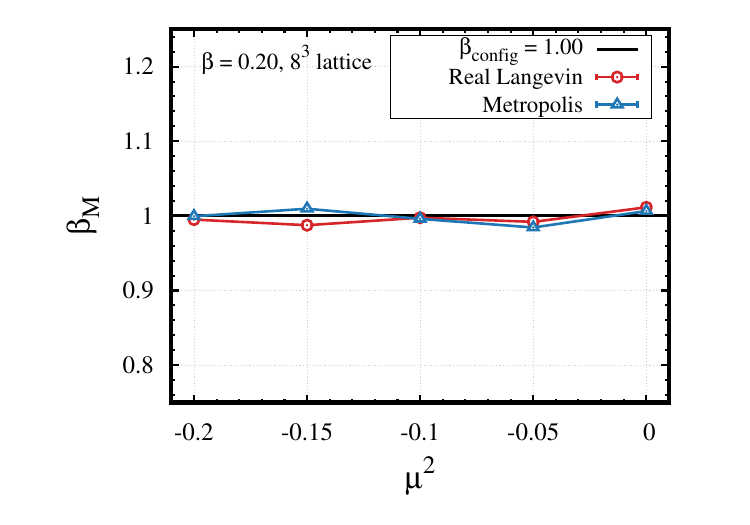} 
\caption{Configurational-temperature estimator $\beta_M$ as a function of $\mu^2$ on an $8^3$ lattice for several values of $\beta$. 
Real Langevin and Metropolis results are shown together. 
The dashed horizontal line denotes the equilibrium expectation $\beta_M = 1$.} 
\label{fig:config_temperature} 
\end{figure}

The configurational-temperature estimator is the principal observable investigated in this work. 
Figure~\ref{fig:config_temperature} shows the measured value of $\beta_M$ as a function of the imaginary chemical potential for the same range of couplings considered above. 
The dashed horizontal line indicates the expected equilibrium value, 
\[ 
\beta_M = 1. 
\] 

For most of the parameter range studied, the configurational-temperature estimator remains close to its expected value. 
Moreover, the measurements obtained from Metropolis Monte Carlo and real Langevin dynamics are generally consistent within statistical uncertainties. 
This agreement is encouraging because the estimator depends directly on the local derivatives of the action and, therefore, provides a more stringent test of the equilibrium sampling than conventional thermodynamic observables. 

We note that the largest deviations from unity happen in the vicinity of the phase transition, where the critical coupling at vanishing chemical potential is approximately $\beta_c \simeq 0.454$~\cite{Campostrini:2000iw}. 
Simulations at $\beta = 0.4$ and $\beta = 0.5$ therefore probe the neighborhood of the critical region. 
Near the transition, the growth of the correlation length and the onset of critical slowing down enhance finite-volume effects and increase the sensitivity of derivative-based observables. 
On the relatively small $8^3$ lattice used in this work, these effects are expected to modify the balance between the gradient and Laplacian contributions entering the configurational-temperature estimator. 

We conclude that the close agreement between the two simulation algorithms implies that the configurational-temperature estimator gives a robust diagnostic of equilibrium sampling. 
The deviations from unity remain modest and can be explained by finite-volume and discretization effects rather than by deficiencies of the underlying simulation algorithms. 
These results establish an important benchmark for future investigations of lattice field theories with complex actions, where independent validation methods are generally unavailable. 

\section{Conclusions} 
\label{sec:conclusions} 

In this work, we investigated the configurational-temperature estimator as a diagnostic tool for lattice field theory simulations using the three-dimensional XY model at an imaginary chemical potential. 
We used Metropolis Monte Carlo and real Langevin dynamics to generate equilibrium configurations since the Euclidean action remains real in this regime; this, in turn, provided a controlled environment for benchmarking the estimator. 

The numerical simulations were first validated through measurements of the action density. 
At weak coupling, the results agree well with the predictions of the strong-coupling expansion. 
At the same time, the close agreement between the Metropolis and Langevin simulations across the full parameter range provides confidence that both algorithms sample the same equilibrium distribution. 

The configurational-temperature estimator was then evaluated over a range of lattice couplings and imaginary chemical potentials. 
For most of the explored parameter space, the estimator remains close to its expected equilibrium value, with consistent results from the two independent simulation algorithms. 
We attribute the modest deviations observed near the critical region to finite-volume effects. 

Our results show that the configurational temperature provides a useful internal consistency check for lattice simulations and complements conventional thermodynamic observables in assessing the quality of equilibrium sampling. 
Establishing the behavior of the estimator in a theory with a real Euclidean action is an important prerequisite for its application to theories with genuinely complex actions. 

The present study, therefore, provides a benchmark for future investigations of lattice field theories at finite density using complex Langevin dynamics, where independent validation methods are generally unavailable and robust diagnostic tools are essential.

\section*{Acknowledgements}

KN would like to thank MITP for the financial support that made the presentation of this work possible. 
AJ was supported in part by the Start-up Research Grant from the University of the Witwatersrand, South Africa. 
The authors also gratefully acknowledge support from the National Institute for Theoretical and Computational Sciences (NITheCS), the Mandelstam Institute for Theoretical Physics (MITP), and the School of Physics at the University of the Witwatersrand.

\bibliographystyle{JHEP}
\bibliography{biblio}

@article{Rugh:1997, 
  author = {Rugh, Hans Henrik}, 
  title = {Dynamical Approach to Temperature}, 
  journal = {Phys. Rev. Lett.}, 
  volume = {78}, 
  number = {5}, 
  pages = {772--774}, 
  year = {1997}, 
  month = {Feb}, 
  eprint = {chao-dyn/9701026}, 
  archivePrefix = {arXiv}, 
  publisher = {American Physical Society}, 
  doi = {10.1103/PhysRevLett.78.772}, 
  url = {https://link.aps.org/doi/10.1103/PhysRevLett.78.772} 
}

@article{Butler:1998, 
  author = {Butler, B. D. and Ayton, Gary and Jepps, Owen G. and Evans, Denis J.}, 
  title = {Configurational temperature: Verification of Monte Carlo simulations}, 
  journal = {The Journal of Chemical Physics}, 
  volume = {109}, 
  number = {16}, 
  pages = {6519--6522}, 
  year = {1998}, 
  month = {Oct}, 
  issn = {0021-9606}, 
  doi = {10.1063/1.477301}, 
  url = {https://doi.org/10.1063/1.477301}
}

@article{Jepps:2000, 
  author = {Jepps, Owen G. and Ayton, Gary and Evans, Denis J.}, 
  title = {Microscopic expressions for the thermodynamic temperature},
  journal = {Phys. Rev. E},
  volume = {62}, 
  number = {4}, 
  pages = {4757--4763}, 
  year = {2000}, 
  month = {Oct}, 
  eprint = {cond-mat/9906423}, 
  archivePrefix = {arXiv}, 
  primaryClass = {cond-mat.stat-mech}, 
  publisher = {American Physical Society}, 
  doi = {10.1103/PhysRevE.62.4757}, 
  url = {https://link.aps.org/doi/10.1103/PhysRevE.62.4757}
}

@article{Dhindsa:2025xfv,
    author = "Dhindsa, Navdeep Singh and Joseph, Anosh and Longia, Vamika",
    title = "{Gradient and Hessian-Based temperature estimator in lattice gauge theories: a diagnostic tool for stability and consistency in numerical simulations}",
    eprint = "2508.05595",
    archivePrefix = "arXiv",
    primaryClass = "hep-lat",
    doi = "10.1007/JHEP10(2025)015",
    journal = "JHEP",
    volume = "10",
    pages = "015",
    year = "2025"
}

@article{Meyer-Ortmanns:1996ioo,
    author = "Meyer-Ortmanns, Hildegard",
    title = "{Phase transitions in quantum chromodynamics}",
    eprint = "hep-lat/9608098",
    archivePrefix = "arXiv",
    reportNumber = "HD-THEP-96-12",
    doi = "10.1103/RevModPhys.68.473",
    journal = "Rev. Mod. Phys.",
    volume = "68",
    pages = "473--598",
    year = "1996"
}

@article{Nagata:2021ugx,
    author = "Nagata, Keitaro",
    title = "{Finite-density lattice QCD and sign problem: Current status and open problems}",
    eprint = "2108.12423",
    archivePrefix = "arXiv",
    primaryClass = "hep-lat",
    doi = "10.1016/j.ppnp.2022.103991",
    journal = "Prog. Part. Nucl. Phys.",
    volume = "127",
    pages = "103991",
    year = "2022"
}

@article{Campostrini:2000iw,
    author = "Campostrini, Massimo and Hasenbusch, Martin and Pelissetto, Andrea and Rossi, Paolo and Vicari, Ettore",
    title = "{Critical behavior of the three-dimensional xy universality class}",
    eprint = "cond-mat/0010360",
    archivePrefix = "arXiv",
    reportNumber = "IFUP-TH-2000-31",
    doi = "10.1103/PhysRevB.63.214503",
    journal = "Phys. Rev. B",
    volume = "63",
    pages = "214503",
    year = "2001"
}

@article{Banerjee:2010kc,
    author = "Banerjee, Debasish and Chandrasekharan, Shailesh",
    title = "{Finite size effects in the presence of a chemical potential: A study in the classical non-linear O(2) sigma-model}",
    eprint = "1001.3648",
    archivePrefix = "arXiv",
    primaryClass = "hep-lat",
    doi = "10.1103/PhysRevD.81.125007",
    journal = "Phys. Rev. D",
    volume = "81",
    pages = "125007",
    year = "2010"
}

@inproceedings{Longia:2026doi,
    author = "Longia, Vamika and Dhindsa, Navdeep Singh and Joseph, Anosh",
    title = "{Configurational Thermometer for Lattice Gauge Theories}",
    booktitle = "{42th International Symposium on Lattice Field Theory}",
    eprint = "2601.17436",
    archivePrefix = "arXiv",
    primaryClass = "hep-lat",
    reportNumber = "TIFR/TH/26-6",
    month = "1",
    year = "2026"
}

@article{Joseph:2025xbn,
    author = "Joseph, Anosh and Kumar, Arpith",
    title = "{Configurational Temperature as a Diagnostic for Complex Langevin Dynamics in the 3D XY Model}",
    eprint = "2509.13314",
    archivePrefix = "arXiv",
    primaryClass = "hep-lat",
    month = "9",
    year = "2025"
}

@article{Joseph:2026hom,
    author = "Joseph, Anosh",
    title = "{Probing Probability Geometry with Schwinger--Dyson Identities: Score Mismatch, Fisher Information, and Configurational Temperature}",
    eprint = "2606.27360",
    archivePrefix = "arXiv",
    primaryClass = "hep-th",
    month = "6",
    year = "2026"
}

@article{Joseph:2026rwh,
    author = "Joseph, Anosh and Mamale, Vinod",
    title = "{Configurational Temperature in Matrix Models and Random Matrix Ensembles}",
    eprint = "2606.28148",
    archivePrefix = "arXiv",
    primaryClass = "hep-th",
    month = "6",
    year = "2026"
}

@inproceedings{Joseph:2026xti,
    author = "Joseph, Anosh and Kumar, Arpith",
    title = "{Thermodynamic Consistency as a Reliability Test for Complex Langevin Simulations}",
    booktitle = "{42th International Symposium on Lattice Field Theory}",
    eprint = "2601.20527",
    archivePrefix = "arXiv",
    primaryClass = "hep-lat",
    month = "1",
    year = "2026"
}

\end{document}